\documentstyle[stwol]{article}

\def\be{\begin{equation}}
\def\ee{\end{equation}}
\def\bea{\begin{eqnarray}}
\def\eea{\end{eqnarray}}

\begin{document}


\title{LARGE SQUARK-MIXING IMPACT ON $H^+$ DECAY IN THE MSSM}

\author{ K. HIDAKA }

\address{Department of Physics, Tokyo Gakugei University, Koganei, 
Tokyo 184, Japan\\ E-mail: hidaka@u-gakugei.ac.jp}


\twocolumn[\maketitle\abstracts{We study the decays of
 the charged Higgs boson $H^+$ within the Minimal Supersymmetric 
Standard Model. We find that the
supersymetric mode $\tilde{t}\bar{\tilde{b}}$ can dominate the $H^+$ decays 
in a wide range of the
model parameters due to the large Yukawa couplings and mixings of 
$\tilde{t}$ and $\tilde{b}$. Compared to the conventional modes 
$\tau^+\nu_\tau$ and $t\bar{b}$, 
this mode has very distinctive signatures. This could have a decisive impact 
on $H^+$ searches at future colliders. We find also that the QCD corrections 
to the $\tilde{t}\bar{\tilde{b}}$ mode are significant, but that they do not 
invalidate our tree-level conclusion above.  
( Invited talk at the 28th International Conference on High Energy
 Physics, Warsaw, Poland, 25-31 July 1996 (to appear in the proceedings
  (World Scientific)); Report-no: TGU-20, hep-ph/9706440 )}]

In the Minimal Supersymmetric Standard Model(MSSM)~\cite{hk} two Higgs doublets
 are necessary, leading to five physical Higgs bosons $h^o,H^o,A^o$ and 
$H^\pm$~\cite{hh}~\cite{gh}. In this article we study the $H^\pm$ decays 
within the MSSM~\cite{ba}~\cite{bb}. 
If all supersymmetric(SUSY) particles are heavy enough, the $H^+$ 
decays dominantly into t\=b; the decays $H^+ \to \tau^+\nu$ and/or $H^+ \to 
W^+ h^o$ are dominant below the t\=b threshold~\cite{hh}. 
If the decays into charginos and neutralinos $H^+ \to 
\tilde{\chi}^+_k \tilde{\chi}^o_l$ are kinematically 
allowed, they can be important in a sizable
region of the MSSM parameters~\cite{dj}. Here we extend these studies by 
including also the SUSY modes $H^+ \to \tilde{t}_i \bar{\tilde{b}}_j$ 
(i,j=1,2). Here $\tilde{t}_i (\tilde{b}_j )$ are the scalar top (scalar bottom)
 mass eigenstates which are mixtures of $\tilde{t}_L$ and $\tilde{t}_R$ 
($\tilde{b}_L$ and $\tilde{b}_R$). 

The lighter stop $\tilde{t}_1$ can be much lighter than the other squarks and 
even lighter than the t quark due to large \~t-mixing being propotional to 
the top Yukawa coupling $h_t$  and the \~t-mixing parameters $A_t$ and $\mu$. 
Similarly the lighter sbottom $\tilde{b}_1$ can also be lighter than the other 
squarks. In the case of large \~t- and \~b-mixings one also expects the 
$H^+ \tilde{t} \tilde{b}$ coupling to be large since it is essentially 
proportional to the Yukawa couplings $h_{t,b}$ and the squark-mixing 
parameters $A_{t,b}$ and $\mu$. Here we show explicitly that the 
mode $\tilde{t}\bar{\tilde{b}}$ can indeed dominate the $H^+$ decay in a wide 
range of the MSSM parameters as is expected by the observation above.

In the MSSM the properties of the charginos $\tilde{\chi}^+_i$ (i=1,2) and 
neutralinos $\tilde{\chi}^o_j$
(j=1,...,4) are completely determined by the parameters M, $\mu$ and 
$\tan\beta = v_2 / v_1$, assuming $M'=(5/3) \tan^2 \theta_W M$. 
Here M($M'$) is the SU(2)(U(1)) gaugino mass, $\mu$ is 
the higgsino mass parameter, and $v_1 (v_2)$ is the vacumm expectation value 
of the Higgs $H^o_1 (H^o_2)$~\cite{hk}. 
Here $m_{\tilde{\chi}^+_1} < m_{\tilde{\chi}^+_2}$ and $m_{\tilde{\chi}^o_1} 
<...<m_{\tilde{\chi}^o_4}$ . To specify the squark 
sector the additional (soft SUSY breaking) parameters $M_{\tilde{Q}}, 
M_{\tilde{U}}, M_{\tilde{D}}$ (for each generation), and 
A(for each flavor) are neccesary. The mass matrix for stops reads~\cite{gh}:
\begin{equation}
   {\cal M}^2_{\tilde{t}}=
      \left(
    	  \begin{array}{cc}
          m_{\tilde{t}_L}^2 & a_t m_t \\
          a_t m_t & m_{\tilde{t}_R}^2
          \end{array}
      \right)\label{eq:a}
\end{equation}
with
\begin{eqnarray}
m_{\tilde{t}_L}^2 &=& M_{\tilde{Q}}^2 + 
                      m_Z^2\cos 2\beta(I^3_t-e_t\sin ^2\theta_W) \nonumber \\
                  & & {} + m_t^2                              \label{eq:b} \\
m_{\tilde{t}_R}^2 &=& M_{\tilde{U}}^2 + 
                      m_Z^2\cos 2\beta e_t\sin^2\theta_W + m_t^2
                                                              \label{eq:c} \\
a_t m_t           &=& -m_t(A_t + \mu\cot\beta).               \label{eq:d}
\end{eqnarray}
Here notice our sign conventions of $\mu$ and A; for $\mu$ we use the sign
convention of Ref.1. For the \~b system analogous formulae hold but 
with $M_{\tilde{U}}^2$ replaced by $M_{\tilde{D}}^2$ in eq.(\ref{eq:c}), and 
instead of eq.(\ref{eq:d})
\begin{equation}
a_b m_b = -m_b(A_b + \mu\tan\beta). \label{eq:e}
\end{equation}
$\tilde{b}_L-\tilde{b}_R$ mixing may also be important for large $A_b,\ \mu$, 
and $\tan\beta$. Analogous formulae hold for the sleptons $\tilde{l}$ and 
$\tilde{\nu}$.

The masses and couplings of the Higgs bosons $H^\pm, H^o, h^o$ and $A^o$, 
including radiative corrections, are fixed by $m_{A^o}, \tan\beta$ , 
$m_t, M_{\tilde{Q}}$, $M_{\tilde{U}}, M_{\tilde{D}}$, 
$A_t, A_b$ and $\mu $~\cite{ba}. 
$H^o (h^o )$ and $A^o$  are the heavier(lighter) CP-even and CP-odd neutral 
Higgs bosons, respectively. For $m_{H^+}$ we take the tree-level relation 
$m_{H^+}^2 = m_{A^o}^2 + m_W^2$, because in all cases considered here 
the radiative corrections to $m_{H^+}$ turn out to be very small. 

In the following, we take for simplicity 
$M_{\tilde{Q}} = M_{\tilde{U}} = M_{\tilde{D}}$  (for the third
generation), $M_{\tilde{L}} = M_{\tilde{Q}}$  ($M_{\tilde{L}}$ being 
a common soft-SUSY-breaking mass of all
sleptons), and $A_t =A_b =A_{\tau}\equiv A$. Thus we have 
$m_{H^+},m_t $,M,$\mu ,\tan\beta ,M_{\tilde{Q}} $, and A as free
parameters of the MSSM. The theoretical and experimental constraints for
the MSSM basic parameters are discribed in Refs.4,7.

We calculate the widths of all important modes of $H^+$  decay: 
$H^+\to t\bar{b}$, $\tau^+\nu$, $c\bar{s}$, $\tilde{t}_i \bar{\tilde{b}}_j$, 
$\tilde{\chi}^+_k \tilde{\chi}^o_l$, $W^+h^o$, $\tilde{l}^+ \tilde{\nu}$. 
Formulae for these widths are found in Ref.2.
As the squarks of the first two generations are supposed to be heavy, these 
decays will be strongly phase-space suppressed. 
In order not to vary too many parameters, 
in the following we fix $m_t =150GeV$ and $\mu =300GeV$, and take the values 
of M and $\tan\beta$ such that $m_{\tilde{\chi}_1^o} \simeq$ 50GeV.

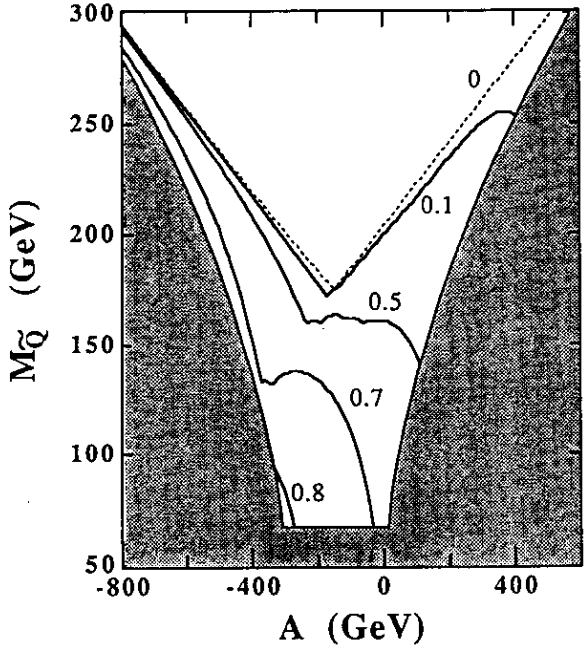
\begin{figure}
\center
\vskip 8.5cm
\caption{Contour lines of $B(\tilde{t}\bar{\tilde{b}})$ in 
the A-$M_{\tilde{Q}}$ plane for ($m_{H^+}$,M,$\mu$,$\tan\beta$)=
(400GeV,120GeV,300GeV,2). The shaded area is excluded by the LEP bound 
$m_{\tilde{\nu}}\stackrel{>}{\sim}45GeV$ and the requirement 
$m_{\tilde{t}_1,\tilde{b}_1,\tilde{l}} > m_{\tilde{\chi}_1^o}$ 
($\simeq$50GeV).}
\label{fig:a}
\end{figure}

In Fig.~\ref{fig:a} the contour lines for the branching ratio 
$B(\tilde{t}\bar{\tilde{b}})$=$\sum_{i,j=1,2} 
B(H^+ \to \tilde{t}_i \bar{\tilde{b}}_j)$ 
are plotted in the A-$M_{\tilde{Q}}$ plane for $m_{H^+}$=400GeV, 
$\tan\beta$=2, M=120GeV. In the plot we have required
$m_{\tilde{t}_1,\tilde{b}_1,\tilde{l}} > m_{\tilde{\chi}_1^o}$ 
($\simeq $50GeV). We see that the branching ratio 
$B(\tilde{t}\bar{\tilde{b}})$ can be
larger than 70\%  in a sizable region. In this region this decay mode is much
more important than the conventional decay modes. 
For large $\tan\beta$ ($\tan\beta$=12) we have obtained a similar result 
to Fig.~\ref{fig:a}~\cite{ba}.

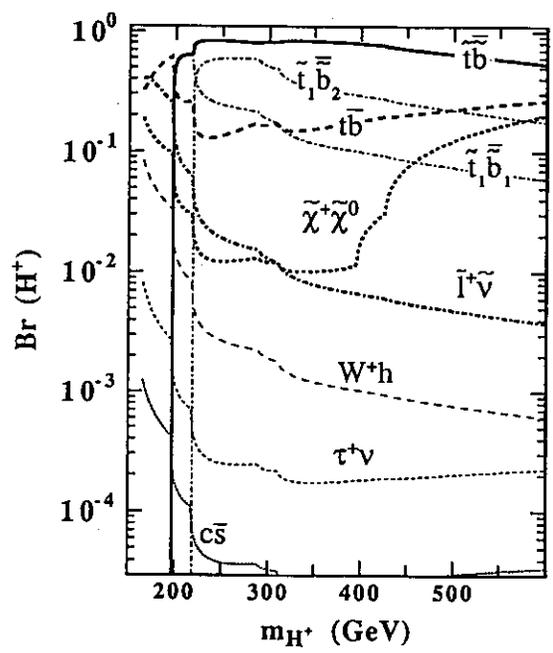
\begin{figure}
\center
\vskip 8.5cm
\caption{The $m_{H^+}$ dependence of all important 
branching ratios of the $H^+$ decay for 
(M, $\mu$, $\tan\beta$, $M_{\tilde{Q}}$, A) = 
(120GeV, 300GeV, 2, 85GeV, -250GeV).
The sum over all mass eigenstates and/or flavours is taken for 
$B(\tilde{t} \bar{\tilde{b}})$, $B(\tilde{\chi}^+ \tilde{\chi}^o)$, and 
$B(\tilde{l}^+ \tilde{\nu})$. $B(\tilde{t}_1 \bar{\tilde{b}}_{1,2})$ 
are also shown separately.}
\label{fig:b}
\end{figure}

In Fig.~\ref{fig:b} we show the $m_{H^+}$ dependence of the important 
branching ratios for $M_{\tilde{Q}}$=85GeV, A= -250GeV, $\tan\beta$=2, 
M=120GeV. In this case we have: $m_{\tilde{t}_1}$=116GeV, 
$m_{\tilde{t}_2}$=209GeV, $m_{\tilde{b}_1}$=81GeV, $m_{\tilde{b}_2}$=102GeV, 
$m_{\tilde{\chi}_1^+}$=94GeV. We see that above the 
$\tilde{t}_1\bar{\tilde{b}}_1$ threshold the $\tilde{t}\bar{\tilde{b}}$ 
mode dominates over the conventional modes t\=b and $\tau^+\nu$. 
For $\tan\beta$=12 we have obtained a similar result 
to Fig.~\ref{fig:b}~\cite{ba}. 

The reason for the dominance of the $\tilde{t}\bar{\tilde{b}}$ mode 
is as follows: The modes t\=b and $\tilde{t}\bar{\tilde{b}}$ 
(whose couplings to $H+$ are essentially $\sim$ $h_t \cos\beta$ + 
$h_b \sin\beta$ and $\sim$ $(A-\mu\tan\beta)$$h_t \cos\beta$ + 
$(A-\mu\cot\beta)$$h_b \sin\beta$, respectively) can be strongly enhanced 
relative to the other modes due to the large Yukawa couplings $h_{t,b}$. 
In addition, the $\tilde{t}\bar{\tilde{b}}$ mode can be strongly enhanced 
relative to the t\=b mode in the case the \~q-mixing parameters A 
and $\mu$ are large. Moreover in this case $\tilde{t}_1$ and 
$\tilde{b}_1$ tend to be light.

   Quite generally, B($\tilde{t}\bar{\tilde{b}}$) depends on the parameters 
$M_{\tilde{Q}}$,A,$m_t,\mu$,$\tan\beta$ and 
more weakly on M. For a given $m_{H^+}$ the strongest dependence is that on 
$M_{\tilde{Q}}$ to which $m_{\tilde{t}}$ and $m_{\tilde{b}}$ are sensitive 
(see Fig.~\ref{fig:a}). B($\tilde{t}\bar{\tilde{b}}$) can be quite large in a 
substantial part of the parameter region kinematically allowed for the 
$\tilde{t}\bar{\tilde{b}}$ mode. We find that the dominance of the 
$\tilde{t}\bar{\tilde{b}}$ mode is fairly insensitive to 
the assumption $M_{\tilde{Q}}$ = $M_{\tilde{L}}$. As seen in 
Fig.~\ref{fig:a} the dependence on A is also strong.
Concerning the assumption $A_t = A_b = A_{\tau}$, we have found no 
significant change of B($\tilde{t}\bar{\tilde{b}}$) as compared to 
Fig.~\ref{fig:b}, when we take $A_{b,\tau}/A_t$ = $\pm 0.5,\pm 1,\pm 2$ 
keeping $A_t$ = A. The dependence of the $H^+ \tilde{t} \bar{\tilde{b}}$ 
couplings (and B($\tilde{t}\bar{\tilde{b}}$)) on the parameters $\mu$, 
$\tan\beta$ and $m_t$ is essentially given by the terms 
$(A-\mu\tan\beta)$$h_t \cos\beta$ and $(A-\mu\cot\beta)$$h_b \sin\beta$ 
as mentioned above, where $h_t \propto m_t$ . Hence {\it the dominance of the 
$\tilde{t}\bar{\tilde{b}}$ mode becomes more pronounced as $m_t$ 
and/or $\mu$ increase.} We also find that B($\tilde{t}\bar{\tilde{b}}$) is 
nearly invariant under ($\mu$,A)$\to$(-$\mu$,-A).

As for the signatures of the $H^+$ dcay, typical $\tilde{t}\bar{\tilde{b}}$ 
signals are shown in Table~\ref{tab:a}. They have to be compared with 
the conventional t\=b signals $H^+ \to t\bar{b}$ $\to (W^+b)\bar{b}$ 
$\to f\bar{f}'b\bar{b}$, i.e. 4 jets(j's) or 2 j's + 1 isolated charged 
lepton ($l^+$) + missing energy-momentum ($p \!\! /$). Note that 
B($\tilde{t}_1 \to c \tilde{\chi}_1^o$) $\simeq$ 1 if $m_{\tilde{t}_1} < 
m_{\tilde{\chi}_1^+,\tilde{l},\tilde{\nu},\tilde{b}_1}$ and 
$m_{\tilde{\chi}_1^o} < m_{\tilde{t}_1}$ 
$< m_{\tilde{\chi}_1^o} + m_{t,W}$ (in cases (a)-(d)), and 
B($\tilde{t}_1 \to c \tilde{\chi}_1^o$) $\simeq$ 0 otherwise 
(in cases (e)-(h))~\cite{ba}. As seen in Table~\ref{tab:a}, the 
$\tilde{t}\bar{\tilde{b}}$ signals have
general features which distinguish them from the t\=b signals: (i) more 
$p \!\! /$ due to the emmision of two LSP's (i.e. $\tilde{\chi}_1^o$'s) 
and hence less energy-momentum of jets and the isolated $l^+$  in case 
of a short decay chain, or (ii) more jets and/or more isolated $l^\pm$'s 
in case of a longer decay chain. 
Moreover, depending on the values of the MSSM parameters, 
the $\tilde{t}\bar{\tilde{b}}$ signals could have 
remarkable features as seen in Table~\ref{tab:a}: (i)the semileptonic 
branching ratio of $H^+$  decay with an isolated $l^+$  can vary 
between $\sim$ 0 (e.g. in case B($H^+ \to$ $\tilde{t}_1\bar{\tilde{b}}_1 \to$ 
$c \tilde{\chi}_1^o \bar{b} \tilde{\chi}_1^o$) $\sim$ 1) and $\sim$ 1 
(e.g. in case B($H^+ \to$ $\tilde{t}_1\bar{\tilde{b}}_1 \to$ 
$(b l^+ \tilde{\nu})$$(\bar{b} \tilde{\chi}_1^o) \to$ $b l^+ 
\nu \tilde{\chi}_1^o$$\bar{b} \tilde{\chi}_1^o$) $\sim$ 1); 
(ii)production of a single "wrong"-sign $l^-$  in $H^+$  decay (e.g. 
in case (d,h)) or same-sign dileptons $l^+ l'^+$ (e.g. in case (h)), 
which could yield same-sign isolated dilepton events $e^+ e^-$ 
(or $\gamma \gamma$) $\to H^+ H^-$ $\to(l^-l'^- or l^+l'^+)$ + j's + 
$p \!\! /$; and so on. The identification of the sign of charged 
leptons and the tagging of b- and c-quark jets, $h^o$, $Z^o$ and $W^\pm$ 
would be very useful in discriminating the 
$\tilde{t}\bar{\tilde{b}}$ signals from 
the t\=b signals as well as in suppressing the background. 
We see that the $\tilde{t}\bar{\tilde{b}}$ signals are very 
different from the conventional t\=b and $\tau^+\nu$ signals. 
If the $\tilde{t}\bar{\tilde{b}}$ mode really dominates the $H^+$ decay, 
it decisively influences the signatures of $H^+$.

\begin{table*}[t]
\caption{The typical $\tilde{t}\bar{\tilde{b}}$ signals of the $H^+$ decay 
in comparison to the conventional t\=b signals. $p \!\! /$, j, $l^\pm$, 
$Z^{(\ast)}$, and $f$ denote missing energy-momentum, jet, isolated charged 
lepton, real(or virtual) $Z^o$ boson, and (q, $l^\pm$, $\nu$), respectively. 
\label{tab:a}}
\vspace{13.0cm}
\end{table*}

We have shown that the SUSY mode $\tilde{t}\bar{\tilde{b}}$ can be the 
most important $H^+$ decay channel in a large allowed region of 
the MSSM parameter space due to large t and b quark Yukawa couplings 
and large $\tilde{t}$- and $\tilde{b}$-mixings. 
The $\tilde{t}\bar{\tilde{b}}$ mode has 
very distinctive signatures as compared to the conventional modes $\tau^+\nu$ 
and t\=b. This could decisively influence the $H^\pm$ search at future 
colliders. We have found that the QCD corrections to the 
$\tilde{t}\bar{\tilde{b}}$ mode are significant,
but that they do not invalidate our tree-level conclusion above~\cite{bb}. 
Finally, we have obtained a conclusion for the $H^o$ and $A^o$ decays 
($H^o, A^o \to \tilde{t}\bar{\tilde{t}}, \tilde{b}\bar{\tilde{b}}$, ...) 
quite similar to one for the $H^+$ decay~\cite{be}.

\section*{Acknowledgments}
This article is based on Refs.4,5.


\section*{References}
\begin{thebibliography}{99}
\bibitem{hk}H. E. Haber and G. L. Kane, Phys. Rep. 117 (1985) 75.

\bibitem{hh}J.F. Gunion, H.E. Haber, G.L. Kane, and S. Dawson, 
The Higgs Hunter's Guide, Addison-Wesley (1990).

\bibitem{gh}J.F. Gunion and H.E. Haber, Nucl.Phys. B272(1986) 1; 
B402(1993)567(E).

\bibitem{ba}A. Bartl, K.Hidaka, Y. Kizukuri, T. Kon and W. Majerotto, 
Phys. Lett. B315(1993)360.

\bibitem{bb}A. Bartl, H. Eberl, K. Hidaka, T. Kon, W. Majerotto and 
Y. Yamada, Phys. Lett. B373(1996)117.

\bibitem{dj}J. F. Gunion and H. E. Haber, Nucl. Phys. B307 (1988) 445;
   A. Djouadi, J. Kalinowski, and P. M. Zerwas, Z. Phys. C57 (1993) 569.

\bibitem{be}A. Bartl, H. Eberl, K. Hidaka, T. Kon, W. Majerotto and 
Y. Yamada, Phys. Lett. B389(1996)538.

\end{thebibliography}
\end{document}